\documentclass[12pt]{article}
\usepackage{amsmath}
\usepackage{amssymb}\usepackage{epsf}
\usepackage{graphicx,color}
\usepackage{eepic}
\usepackage{latexsym}
\usepackage{epsfig}
\usepackage[colorlinks]{hyperref}

\let\Large=\normalsize

\newcommand{\rcite}[1]{{\cite{#1}}}

\newcommand{\rbibitem}[1]{\bibitem{#1}}
\newcommand{\tr}{\mbox{\rm tr}}
\newcommand{\be}{\begin{equation}}
\newcommand{\ee}{\end{equation}}
\newcommand{\ba}{\begin{eqnarray}}
\newcommand{\ea}{\end{eqnarray}}
\newcommand{\dis}{\displaystyle}

 \def\tr {\, \mbox{tr} \,}

\usepackage{epsf}
\usepackage{graphicx}
\usepackage{eepic}
\usepackage{latexsym}

\def\sl{\llap{$/$}}%
\begin{document}

\begin{center}

{\Large\bf  On the   scalar nonet in the extended Nambu Jona-Lasinio
model }
\\[10mm]
{\sc M.~X.~Su, L.~Y.~Xiao and H.~Q.~Zheng}
\\[2mm]
{\it  Department of Physics, Peking University, Beijing 100871,
P.~R.~China}
\\[5mm]
\today

\end{center}

\vskip 1cm
\begin{abstract}
We discuss the lightest scalar resonances, $f_0(600)$,
$\kappa(800)$, $a_0(980)$ and $f_0(980)$ in the extended Nambu
Jona-Lasinio model. We find that the model parameters can be tuned,
but unnaturally,  to accommodate those scalars except the
$f_0(980)$.  We also discuss problems encountered in the K Matrix
unitarization approximation by using $N_c$ counting technique.
\end{abstract}
\section{Introduction}

The original model of Nambu and Jona-Lasinio~\rcite{NJL} (NJL)was
proposed as a dynamical model of the strong interactions between
nucleons and pions,   before the invention of QCD. In this model
pions appear as the massless composite bosons associated with the
dynamical spontaneous breakdown of the chiral symmetry of the
initial lagrangian. Even after the invention of QCD, the NJL model
or the extended NJL (ENJL) model still serves as a useful tool
widely discussed in the literature when discussing low energy strong
interaction physics at the quark-level, starting from last seventies
and eighties~\rcite{QuarkNJL,ENJL}. More recent extensive reviews
can be found in~\rcite{review}.

The ENJL  model provides a natural extension and hence is considered
more general than the linear sigma model or the model in which the
$\rho$ meson takes the role of a massive gauge boson of the isospin
symmetry. The ENJL model attempts to provide a unified description
to both the scalar sector as well as the vector sector, in a chiral
symmetric way. However, there have been controversies for a long
time on the spectrum of the lowest lying scalar nonet in strong
interactions. This situation is also reflected in the early studies
on ENJL models. In Ref.~\rcite{Volkov1}, the ENJL model is used to
study the lightest scalar nonet, and it is found that, $m_a=500$MeV,
$m_{K_*}=840$MeV, $m_{\sigma_0}\simeq 760$MeV,
$m_{\sigma_8}=950$MeV. The mass of $a_0(980)$ meson (which is
degenerate to the SU(2) $\sigma$ in Ref.~\cite{Volkov1}) could not
be explained by the NJL model itself, and the mass difference is
ascribed to a possible $qq\bar q\bar q$ content in the physical
$a_0$. In Ref.~\rcite{Dmi} the role of QCD $U_A(1)$ anomaly is
considered and a sum rule is obtained between the mass of scalars
and the pseudoscalars in the NJL model via 't Hooft's instanton
interaction~\rcite{thooft1}. The instanton effect will break the
degeneracy between the octet and the singlet, which lifts the former
and suppresses the latter. Starting with a `bare' quarkonium mass of
1100MeV in Ref.~\rcite{Dmi}, $U_A(1)$ splits the $0^{+}$ nonet into
a singlet $\thicksim 1000$MeV and an octet $\thicksim 1300$MeV.
Broken $SU_f(3)$  further splits the masses so that one gets:
$m_{a_0}=1320$MeV, $m_{\sigma_8}=1590$MeV, $m_{\sigma_0}=1000$MeV,
$m_{K^*_0}=1430$MeV. Apparently this assignment is not for the
lightest scalars since it contains a heavy $a_0$. The above work did
not include $\kappa$ (or $K^*_0(800)$) in their lightest scalar
nonet. This situation is improved by Volkov $et$
$al$~\rcite{Volkov2} who discussed the NJL model with the 't Hooft
interaction (which splits the mass between $a_0$ and $\sigma$) and
find $m_{a_0}=810$MeV, $m_\sigma=550$MeV, $m_{\sigma'}=1130$MeV,
$m_{K_0^*}=960$MeV. Nevertheless there is an apparent problem with
these results, that is for a small $\sigma$ mass around 550MeV (and
also a small mass for the $\kappa$), it does not possess enough
phase space to develop a large width for the $\sigma$ as is revealed
by recent determinations. More recent work~\rcite{Osipov1,Osipov2}
observed that there is growing evidence that $a_0(980)$,
$K_0^*(800)$ or $\kappa$, as well as $f_0(600)$ or $\sigma$ and
$f_0(980)$, are members of the low-lying scalar nonet. Nevertheless
Ref.~\cite{Osipov1} suffers from the similar problem as in
Ref.~\cite{Volkov2}, it also gives a rather small mass of $\sigma$
with which it is difficult to explain the large width of the sigma
simultaneously.

On the other side, progress has been made in recent few years,
demonstrating the existence of the light and broad
 $\sigma$ (or $f_0(600)$) and $\kappa$  resonances~\rcite{PDG}.
 The pole locations
of $\sigma$ and $\kappa$ are determined using dispersive
approaches~\cite{Leutwyler}--\cite{Zheng3}. This new information on
the pole locations urges and enables us to watch more carefully the
dynamics with respect to the lightest scalars, within the scheme of
the extended NJL model, which is the purpose of this paper.

The basic idea of the present paper is outlined already in
Ref.~\rcite{xiao06}, where we pointed out that in order to
understand
 correctly the mass relations among lightest scalars one has to take into
account the additional information provided by the widths of these
scalars,  ranging from  a few tens MeV to a few hundred MeV.
Especially  when there appears a large width, since it is an
unambiguous signal for strong interactions in the given channel, the
bare mass spectrum at tree level has to be strongly distorted. A
certain unitarization procedure is necessary to explore the relation
between the pole mass parameters and the bare mass parameters put in
the lagrangian. For example it is suggested in Ref.~\rcite{xiao06}
that a $\sigma$ pole locates at $\sqrt{z}=m- i\Gamma/2=470- 285i$
MeV corresponds to a bare mass $M_\sigma\simeq 930 MeV$ with some
uncertainties.

This paper is devoted to the study on the lightest scalar resonances
within the ENJL model. Our aim is to explore whether one can explain
within the model, at least at qualitative level, the masses and
widths of $\sigma(600)$, $K_0^*(800)$, $a_0(980)$ and $f_0(980)$
simultaneously. We find that the model encounters serious
difficulties for its own reason, though it can be finely tuned,
 unnaturally,  to explain the masses and widths of
$f_0(600)$, $K_0^*(800)$ and $a_0(980)$. However in such a case the
ENJL model can no longer  explain the vector meson spectrum.
Furthermore, if we only focus on the scalar sector and ignore the
problem with vector meson mass spectrum, there still remains a
problem that the $f_0(980)$ is not possible to be described as a
member of the scalar octet: it has a too small mass.

This paper is organized as following: section 1 is the
introduction, in section 2, we make a short review on the ENJL
model, especially those materials being used in this paper. In
section 3, we reconstruct the mass relations of scalar mesons  and
also discuss the tree-level decay widths of scalars. In section 4,
using $K$ matrix method we construct a unitarized scattering
amplitude and find pole locations in each channel, numerical
results are listed and discussed. We also discuss the $N_c$
dependence of pole trajectories. Our discussion is also slightly
generalized by including the unitarization approximation of the
more general resonance chiral theory. In section 5, we draw our
conclusions on the nature of light scalars, based on our study on
the $K$ matrix unitarization of the ENJL model amplitude.
\section{The ENJL model}
This section reviews how to derive an effective meson chiral
lagrangian, involving both scalars and vector mesons, from a four
fermi interaction. Combining with 't Hooft's interaction lagrangian
the ENJL model provides the basic tool for our study. The method
introduced in this section is standard~\cite{QuarkNJL,ENJL,review}.
\subsection{Bosonization of the ENJL model}
We start from the four quark interactions
\begin{eqnarray}
\label{ENJL} {\cal L}_{\rm QCD} &\rightarrow &
\sum_i\bar q(i\partial\sl-{\cal M})q + {\cal L}_{\rm NJL}^{\rm S,P}
+ {\cal L}_{\rm
NJL}^{\rm V,A} + {\cal O}\left(1/\Lambda_\chi^4\right),\\
{\rm with}\hspace*{1.5cm} {\cal L}_{\rm NJL}^{\rm S,P}&=& \frac{\dis
8\pi^2 G_S }{\dis N_c \Lambda_\chi^2} \, {\dis \sum_{i,j}}
\left(\overline{q}^i_R
q^j_L\right) \left(\overline{q}^j_L q^i_R\right) \nonumber\\
{\rm and}\hspace*{1.5cm} {\cal L}_{\rm NJL}^{\rm V,A}&=& -\frac{\dis
8\pi^2 G_V}{\dis N_c \Lambda_\chi^2}\, {\dis \sum_{i,j}} \left[
\left(\overline{q}^i_L \gamma^\mu q^j_L\right)
\left(\overline{q}^j_L \gamma_\mu q^i_L\right) + \left( L
\rightarrow R \right) \right] \ ,\nonumber
\end{eqnarray}
where $i,j$ are flavor indices, $\Psi_{R,L} \equiv (1/2) \left(1 \pm
\gamma_5\right) \Psi$ and the couplings $G_S$ and $G_V$ are
dimensionless quantities. We adopt the same symbols and definitions
as in the third reference of Ref.~\cite{review}. We  introduce three
complex $3\times 3$ auxiliary field matrices $M(x)$, $L_{\mu}(x)$
and $R_{\mu}(x)$, which under the chiral group $G=SU_L(3)\times
SU_R(3)$ transform as
\begin{eqnarray}
\label{auxiliary} M &\to& g_R M g_L^{\dagger}, \nonumber \\
L_{\mu}\to g_L
L_{\mu} g_L^{\dagger}\,\,\,&{\hbox{and}}&\,\,\,R_{\mu}\to g_R
 R_{\mu} g_R^{\dagger}.
\end{eqnarray}
By polar decomposition \be M = U \tilde H = \xi H \xi, \ee with $U$
unitary, $\tilde H$ (and $H$) hermitian and
 \be \xi(\Phi) \to g_R \xi(\Phi)
h^{\dagger}(\Phi,g_{L,R}) = h(\Phi,g_{L,R}) \xi(\Phi) g_L^{\dagger},
\ee where $\xi(\Phi) \xi(\Phi) = U$.
 From the transformation laws of $M$ and $\xi$, it follows that $H$ transforms homogeneously, i.e.,
\be H \to h(\Phi,g_{L,R}) H h^{\dagger}(\Phi,g_{L,R}). \ee We can
reconstruct the vector fields
 \be W_{\mu}^{\pm} =
\xi L_{\mu} \xi^{\dagger} \pm \xi^{\dagger} R_{\mu} \xi. \ee
 The
transformation properties is
 \be W_{\mu}^{\pm} \to
h(\Phi,g) W_{\mu}^{\pm} h^{\dagger}(\Phi,g)\ .
 \ee
 After some
deduction, one obtains in the Euclidean space the effective action
 $\Gamma_{eff}(M_Q,\xi,\sigma,W_{\mu}^{\pm};v,a,s,p)$ in terms of the
new auxiliary field variables and in the presence of the external
field sources $v_{\mu}$, $a_{\mu}$, $s$ and $p$~\cite{review},
\begin{displaymath}
e^{\Gamma_{eff}(M_Q,\xi,\sigma,W_{\mu}^{\pm};v,a,s,p)} =
\end{displaymath}
\begin{displaymath}
\hbox{exp}\left( - \int d^4 x \left\{ {N_c \Lambda_{\chi}^2 \over
8\pi^2 G_S(\Lambda_{\chi})}tr H^2 + {N_c \Lambda_{\chi}^2 \over
16\pi^2 G_V(\Lambda_{\chi})}{1 \over 4}tr(W_{\mu}^+
 W_{\mu}^+ +
W_{\mu}^- W_{\mu}^-)\right\}\right)\times
\end{displaymath}
\be 
\int {\cal D}\bar{Q} {\cal D}Q \hbox{exp} \int d^4 x \bar{Q}{\cal
D}_E Q, \ee
 where
${\cal D}_E$ denotes the Euclidean Dirac operator
\be\label{99} {\cal D}_E = \gamma_{\mu} \nabla_{\mu} - {1 \over
2}(\Sigma - \gamma_5 \Delta) - H(x) \ee
 with
  $\nabla_{\mu}$,
   the
covariant derivative
 \be \nabla_{\mu}=\partial_{\mu} +
 \Gamma_{\mu} - {i \over 2} \gamma_5 (\xi_{\mu} -
W_{\mu}^{(-)}) - {i \over 2} W_{\mu}^{(+)} \ee  and
 \be
\Sigma = \xi^{\dagger} \cal{M} \xi^{\dagger} + \xi \cal{M}^{\dagger}
\xi \ee \be \Delta = \xi^{\dagger} \cal{M} \xi^{\dagger} - \xi
\cal{M}^{\dagger}
 \xi\ .
 \ee
The quantities  $\Gamma_{\mu}$ and $\xi_{\mu}$ are those
 \be \Gamma_{\mu} = {1 \over
2}\{\xi^{\dagger}[\partial_{\mu}-i(v_{\mu}+a_{\mu})]\xi +
\xi[\partial_{\mu}-i(v_{\mu}-a_{\mu})]\xi^{\dagger}\}\ , \ee
\noindent and
 \be \xi_{\mu} =
i\{\xi^{\dagger}[\partial_{\mu}-i(v_{\mu}+a_{\mu})]\xi -
\xi[\partial_{\mu}-i(v_{\mu}-a_{\mu})]\xi^{\dagger}\} =
i\xi^{\dagger} D_{\mu} U \xi^{\dagger}=\xi_{\mu}^{\dagger}\ .
 \ee
 The
effective action is in the basis of constituent chiral quark fields
$Q$,
 \be Q_L = \xi q_L\,,\,\bar Q_L = \bar q_L
\xi^{\dagger}\,; \,Q_R = \xi^{\dagger} q_R \,,\,\bar Q_R = \bar q_R
\xi\ . \ee
With
 \be \hbox{exp } \int {\cal D}\bar{Q}
{\cal D}Q \hbox{exp} \int d^4 x \bar{Q}{\cal D}_E Q =
 \hbox{det} {\cal D}_E,
\ee
 we can get the effective action
\begin{eqnarray}
\label{effectiveaction}
&&\Gamma_{eff}(M_Q,\xi,\sigma,W_{\mu}^{\pm};v,a,s,p)\nonumber \\
&&=-\int d^4 x \left\{ {N_c \Lambda_{\chi}^2 \over 8\pi^2 G_S}tr H^2
+ {N_c \Lambda_{\chi}^2 \over 16\pi^2 G_V}{1 \over 4}tr(W_{\mu}^+
 W_{\mu}^+ +
W_{\mu}^- W_{\mu}^-)\right\}\nonumber \\
&&+\ln( \hbox{det} |{\cal D}_E|).
\end{eqnarray}
Using proper time regularization and heat-kernel expansion
method~\rcite{heatkernel}, we get an effective Lagrangian of meson
fields from the ENJL model. One can also use other regularization
method~\rcite{Wuyueliang} to get similar effective lagrangian.
\subsection{Gap Equation and the 't Hooft Interaction}
Here, we are looking for translational invariant solutions which
minimize the effective action, i.e.,
\begin{equation}
\label{minimize} {\delta \Gamma_{eff}(H,...) \over \delta
H}\vert_{L_{\mu}=R_{\mu}=0,\xi=1,H=<H>; v_{\mu}=a_{\mu}=s=p=0} = 0,
\end{equation}
where $<H> = \hbox{diag} (M_u,M_d,M_s)$. The minimum is reached when
 all the eigenvalues
of $<H>$ are equal, i.e.,
 \be\label{64} <H> = M_Q 1 \ee
 and the
minimum condition leads to the so called gap equation
\begin{equation}
\label{gapeqn1} \int d^4x\,\mathrm{Tr}(x\vert D_E^{-1}\vert x)\vert_
 {L_{\mu}=R_{\mu}=0,\xi=1,H=M_Q;v_{\mu}=a_{\mu}=s=p=0}
= - 4 M_Q {N_c \Lambda_{\chi}^2 \over 16\pi^2
G_S(\Lambda_{\chi})}\int d^4 x,
\end{equation}
where  $M_Q$ is the constituent quark mass. From Eq.~(\ref{gapeqn1})
one further gets,
\begin{eqnarray}
\label{gapeqn2} M_Q  =  { G_S \over \Lambda_{\chi}^2} \Gamma(-1,
\frac{M_Q^2}{\Lambda_{\chi}^2}) M_Q^3,
\end{eqnarray}
where $\Gamma(-1,x)$ denotes the incomplete gamma function
\be\label{gammafunction}
\Gamma(n-2,x=\frac{M_Q^2}{\Lambda_{\chi}^2}) = \int_{M_Q^2 /
\Lambda_{\chi}^2}^{\infty}{dz \over z}e^{-z} z^{n-2} ;
 \,\,\,\,\, n=1,2,3,...\ .
\ee The Eq.~(\ref{gapeqn2}) is obtained using proper time
regularization method used in this paper.
The  gap equation (\ref{gapeqn2}) is obtained without introducing
the current quark masses. We can introduce the current quark mass
through the external source field $s= \hbox{diag} (m_u,m_d,m_s)$,
$m_{u, d, s}$ is the current quark mass. Unlike the method used in
Ref.~\rcite{Osipov2} we just use the gap equation without explicit
$SU_f(3)$ breaking to avoid the complicated calculation in
heat-kernel expansion. We need to shift the the SU(3) singlet field
$\sigma_0$ and the octet filed $\sigma_8$ again in the broken phase
to get the physical fields with zero vacuum expectation values in
the effective lagrangian.

The next step is to add the 't Hooft interaction~\rcite{thooft1},
$\beta (\hbox{det}H + \hbox{det}H^\dagger)$, where $\beta$ is a
constant characterizing the strength of the anomaly contribution. We
get the modified gap equation,
\begin{eqnarray}
\label{minimize1} &\displaystyle{{\delta \Gamma_{eff}'(H,...) \over
\delta H}\vert_{H=<H>}}& = {\delta \Gamma_{eff}(H,...) \over \delta
H}\vert_{H=<H>} + \beta{\delta (\hbox{det}H + \hbox{det}H^\dagger)
\over \delta
H}\nonumber \\
&\Rightarrow&
\frac{\Lambda_\chi^2}{G_S}-\Gamma(-1,x)M_Q^2=\frac{8\pi^2\beta
M_Q}{N_C}.
\end{eqnarray}

\subsection{The Effective Lagrangian and its couplings}

In the ENJL model, we have six input parameters:
\begin{equation}
\label{inputparameters} G_S,\,\ G_V,\,\ \Lambda_x,\,\ m_q,\,\
m_s,\,\ \hbox{and}\ \,\ \beta\ .
\end{equation}
The gap equation
\begin{equation}
\label{gapequation}
\frac{\Lambda_\chi^2}{G_S}-\Gamma(-1,x)M_Q^2=\frac{8\pi^2\beta
M_Q}{N_C}
\end{equation}
introduces a constituent chiral quark mass parameter $M_Q$, and the
ratio
\begin{equation}
\label{ratio-x} x=\frac{M_Q^2}{\Lambda_x^2}.
\end{equation}
\hspace{0cm}We can replace the parameters $G_S$, $G_V$ and
$\Lambda_x$ with $x$, $M_Q$, and
\begin{equation} \label{gA} g_A = {1 \over 1 + 4 G_V x
\Gamma(0,x)}\ ,
\end{equation}
characterizing the $\pi$ -- $A_1$ mixing. In the limit of $G_V\to 0
~(g_A\to 1)$, the ENJL model goes back to the NJL model.

The effective Lagrangian can be written down in the form:
\begin{eqnarray}\label{Leff4} {\cal L}_{eff} & = &  \frac{1}{4}
f_{\pi}^2\, \left[ \tr \left(D_{\mu}UD^{\mu}U^{\dagger}\right) + \tr
\left( \chi
  U^{\dagger}+U^{\dagger} \chi \right)\right]  \nonumber \\ & &
-\frac{1}{4}\tr\,[V_{\mu\nu}V^{\mu\nu}-2M_{V}^2 V_{\mu}V^{\mu}]
 \nonumber \\ & &  -\frac{1}{4}\tr\,[A_{\mu\nu}A^{\mu\nu}-2M_{A}^2 A_{\mu}A^{\mu}]
\nonumber
\\ & &  -\frac{1}{2\sqrt{2}}\left[f_{V}\tr\left(
V_{\mu\nu}f_{+}^{\mu\nu}\right)+i g_{V}\tr\left(
V_{\mu\nu}[\xi^{\mu}, \xi^{\nu}]\right)+f_{A}\tr\left(
A_{\mu\nu}f_{-}^{\mu\nu}\right)\right] \nonumber \\
& & +\cal{L}_S\nonumber
\\ & &
   + \tilde{L}_1 \left(\tr\, D_{\mu}U^{\dagger} D^{\mu}U\right)^2+
\tilde{L}_2 \tr\left( D_{\mu} U^{\dagger} D_{\nu}U\tr\,
D^{\mu}U^{\dagger}D^{\nu}U\right) \nonumber
\\ & &  +\tilde{L}_3
\tr\left( D_{\mu}U^{\dagger} D^{\mu}UD_{\nu}U^{\dagger} D^{\nu}U
\right) \nonumber \\ & & + \tilde{L}_5\tr\left[D_{\mu}U^{\dagger}
D^{\mu}U \left(\chi^{\dagger}U+U^{\dagger}\chi\right)\right]
+\tilde{L}_{7}\left[\tr (U\chi^{\dag}-\chi U^{\dag})\right]^2
\nonumber
\\ & &
 +
\tilde{L}_8\tr\left(\chi^{\dagger}U \chi^{\dagger}U +\chi
U^{\dagger}\chi U^{\dagger}\right) +
\tilde{H}_{2}\tr\left(\chi^{\dagger}\chi\right)\, , \end{eqnarray}
where
\begin{eqnarray}
\label{rcts1} \mathcal{L}_S&=& {1 \over 2} \tr\left(d_{\mu}S
d^{\mu}S - M_S^2 S^2\right) + c_m \tr\left(S \chi^+\right) + c_d
\tr\left(S \xi_{\mu}\xi^{\mu}\right)+...
\end{eqnarray}
 The relevant coupling constants in above lagrangians
 are listed in the Appendix A. We do not integrate
out the heavy resonances (vectors, axial-vectors and scalars) to get
the low energy constants $L_i$, therefore we use $\tilde{L}_i$ to
distinguish them from the LECs in chiral perturbation
theory~\rcite{CHPT}.

Attempts have been made in expressing the low energy constants in
terms of QCD operators~\rcite{Wang}. Nevertheless most reliable
estimates and determinations at this stage are from phenomenological
studies~\rcite{ecker}. The couplings $f_V$ and $g_V$ can be then
determined from the decay
 $\rho^0 \to e^+ e^-$
and $\rho \to \pi \pi$ respectively, with the result \[ \vert f_V
\vert = 0.20\,\,\,{\hbox{and}}\,\,\,\vert g_V \vert =
0.090.\nonumber \] The decay $a_1 \to \pi \gamma$ fixes the coupling
\[\vert f_A \vert = 0.097 \pm 0.022.\]
For the scalar couplings $c_m$ and $c_d$, there exists controversy
due to the lowest scalar multiplet. One can take the scalar
multiplet including $a_0(980)$ as the lightest scalar nonet as in
Ref.~\rcite{ecker}. Using the $a_0 \to \eta \pi$ decay width and
assuming the scalar saturation of $L_5$ and $L_8$ to determine $c_m$
and $c_d$, in this way one computes with $M_S=M_{a_0}=983$ MeV,
\begin{eqnarray}
|c_d|&=&32\,\hbox{MeV},\nonumber\\
|c_m|&=&42\,\hbox{MeV},\nonumber\\ \hbox{and}\,\,\,c_d c_m &>&0.
\end{eqnarray}
Alternately, the authors of Ref.~\rcite{Pich} consider the scalar
multiple to be around $1.2\sim1.4$ GeV as the lightest scalars in
the $L_S$ and gives the value
 $c_d=c_m\sim f_\pi/2$.

The coupling constants, $c_m$, $c_d$, $f_V$, $f_A$ and $G_V$ have
been given from the ENJL model in Ref.~\rcite{Bijnens}. As will be
shown later, the most important parameter appearing in this paper is
the axial vector coupling $g_A$. The preferred value of $g_A$ is
found to be around 0.6 in Ref.~\rcite{Bijnens}.  In
Ref.~\rcite{Peris}, an estimation gives $g_A=\frac{1}{2}$ under some
additional theoretical constraints.

\section{Scalar Mass Spectrum and Decays of Scalar Mesons}
\subsection{Scalar Mass Spectrum}
The scalar nonet is denoted as the following,
 \be
  S(x)=
\left(
\begin{array}{ccc}
{\frac{a^{0}}{\sqrt{2}}}+{\frac{\sigma _{0}}{\sqrt{3}}+}\frac{\sigma _{8}}{%
\sqrt{6}} & a^{+} & \kappa ^{+} \\
a^{-} & -{\frac{a^{0}}{\sqrt{2}}+}\frac{\sigma
_{0}}{\sqrt{3}}+{\frac{\sigma
_{8}}{\sqrt{6}}} & \kappa ^{0} \\
\kappa ^{-} & \overline{\kappa }^{0} & \frac{\sigma _{0}}{\sqrt{3}}-\sqrt{%
\frac{2}{3}}\sigma _{8}%
\end{array}%
\right)\ .
 \ee
 Mass relations for the
scalar nonet can be extracted from the effective lagrangian. Firstly
for the charge and flavor neutral scalars there is a mixing term:
\be \label{non-diagonal}{\cal
L}_{mixing}=M_{00}^2\sigma_0^2+M_{08}^2\sigma_0\sigma_8+M_{88}^2\sigma_8^2\
. \ee One finds
\begin{eqnarray} \label{sclar masses}
   &&M_{00}^2=\frac{1}{3}(2M_\kappa^2+M_{a_0}^2)-g_A M_{th}^2 , \nonumber \\
   &&M_{88}^2=\frac{1}{3}(4M_\kappa^2-M_{a_0}^2)\nonumber ,\\
   &&M_{08}^2=-\frac{4\sqrt{2}}{3}(M_\kappa^2-M_{a_0}^2),
\end{eqnarray}
where
\begin{equation} \label{mthooft}
M_{th}^2=m_\eta^2+m_{\eta'}^2-2m_K^2=4\beta
M_Q^3(\frac{1}{2f_K^2-f_\pi^2}+\frac{2}{f_\pi^2})\ ,
\end{equation}
and
\begin{eqnarray}\label{scalar res}
   &&M_{a_0}^2 \simeq  3g_A m_\pi^2+4M_Q^2+\frac{2}{3}g_A M_{th}^2,\nonumber\\
   &&M_\kappa^2\simeq  3g_A m_K^2+4M_Q^2+\frac{2}{3}g_A M_{th}^2,
\label{a kappa masses}
\end{eqnarray}
and $M_Q$ is the constituent quark mass. The Eq.~(\ref{scalar res})
is only exact in the leading order of cutoff dependence. After
diagonalizing Eq.~(\ref{non-diagonal})  one gets the masses for mass
eigenstates and the mixing angle,
\begin{eqnarray}
     &&M^2_\sigma=\frac{1}{2}\left[2M_\kappa^2-g_A M_{th}^2-
                           \sqrt{(M_{00}^2-M_{88}^2)^2+(M_{08}^2)^2}\right], \nonumber \\
   &&M^2_{\sigma'}=\frac{1}{2}\left[2M_\kappa^2-g_A M_{th}^2+
                           \sqrt{(M_{00}^2-M_{88}^2)^2+(M_{08}^2)^2}\right],\nonumber \\
  &&\tan 2\theta ={\frac{4\sqrt{2}}{3}(M_\kappa^2-M_{a_0}^2)\over
  \frac{2}{3}(M_\kappa^2-M_{a_0}^2)+g_A M_{th}^2}\ .
\label{sigma mass}
\end{eqnarray}
 From (\ref{a kappa masses}) and (\ref{sigma mass}), one gets
immediately two sum rules:
\begin{eqnarray}
\label{sumrule1}
 &&2M^2_\kappa -
 M^2_\sigma-M^2_{\sigma'}\simeq g_A M^2_{th}  ,     \\
\label{sumrule2}
 &&M^2_\kappa-M^2_a \simeq 3g_A(m^2_K-m^2_\pi).
\end{eqnarray}
The first sum rule Eq.~(\ref{sumrule1}) has been obtained in
Ref.~\rcite{Dmi}. The second sum rule Eq.~(\ref{sumrule2}) is in
qualitative agreement with the results given in
Ref.~\rcite{Osipov1}. They are the consequence of $U_A(1)$ breaking
in the ENJL model combined with linear $SU(3)$ symmetry breaking
terms. If we do not include the anomaly contribution in the scalar
mass spectrum, just setting $M_{th}^2=0$ in (\ref{a kappa masses}),
we can find the mixing angle $\theta
=(\arctan2\sqrt{2})/2=\theta_{id}\simeq35.26^\circ$. Then the
$\sigma$ is a pure non-strange state, and the $\sigma'$ is purely
strange. The scalar masses become
\begin{eqnarray}
\label{idealmixingscalarmass}
   &&M_{a_0}^2 =  M^2_\sigma=3g_A m_\pi^2+4M_Q^2\ ,\nonumber\\
   &&M_\kappa^2=  3g_A m_K^2+4M_Q^2\ ,\nonumber\\
   &&M^2_{\sigma'}=2M^2_\kappa-m^2_\pi\ .
\end{eqnarray}

Before jumping into more detailed numerical calculations, we can
make some simple estimates and discussions at qualitative level with
Eqs.~(\ref{sumrule1}) and (\ref{sumrule2}). The first important
thing to notice is that, as already emphasized in
Ref.~\cite{xiao06}, the scalar masses appeared in
Eqs.~(\ref{sumrule1}) and (\ref{sumrule2}) are only `bare' mass
parameters appeared in the lagrangian, which, when the interaction
becomes strong, can be totally different from the pole mass
positions. The large width  of $\sigma$ (or $\kappa$) is an
unambiguous signal for a strong $\sigma\pi\pi$ (or $\kappa K\pi$)
interactions. The large widths are quite often ignored when
discussing the mass spectrum in the literature. It is often
attempted to set up SU(3) mass relations among pole mass parameters
$m$. However, a light $\sigma$ with a mass around 500MeV as a bare
parameter appeared in the lagrangian can hardly produce a large
width, in any model calculations. On the other side,
 the parameter $M$ for $\sigma$ or $\kappa$ can be
estimated to be $M_\sigma\simeq 930$MeV and $M_\kappa\simeq 1380$MeV
with sizable error bars~\cite{xiao06}.
Qualitatively speaking the stronger the resonance couples to the
$\pi\pi$ continuum, the larger the deviation is between $m$ and $M$.
Instead of comparing different $m$, one should firstly examine  the
relations between different ``bare" mass parameters, $M$. Since the
former quantities associated with large widths are severely
distorted by the strong couplings to the pseudo-goldstone pairs, it
is not suitable to use them to discuss the SU(3) mass relations. For
example, we have
 \be\label{rel1}
m_\sigma<m_\kappa<m_{a_0}\ ,
 \ee
  but actually the mass relation should
be read as
 \be\label{rel2}
  1GeV\simeq M_\sigma  \lesssim M_{a_0}<M_\kappa\ .
 \ee
The mass of $\sigma'$ is unavoidably large, with or without anomaly
contributions. If we include the contribution of the 't Hooft
interaction, taking for example $M_\kappa=1.2GeV$, $M_\sigma \simeq
1GeV$, $g_A=0.6$, and $M_{th}^2=0.72GeV^2$, we   get
$M_{a_0}=1.02GeV$, $M_{\sigma'}=1.2GeV$, $\theta \simeq 23.59^\circ
$, 
and
\begin{equation}
\label{scalar1}
     {\sigma \choose \sigma'} \simeq \left( \begin{array}{cc}
                               0.98 & -0.20 \\
                               0.20 & 0.98
                               \end{array}
                        \right)
     {\sigma_{ns} \choose -\sigma_s}
\end{equation}
from Eqs.~(\ref{a kappa masses})-(\ref{sumrule2}). If $g_A$ grows
larger, the mass of $\sigma'$ will be  heavier.
If we neglect the contribution of 't Hooft interaction, and taking
for example $M_a=M_\sigma \simeq 1GeV$ and $g_A=0.6$, we   get
$M_\kappa \simeq 1.19 GeV, M_{\sigma'} \simeq 1.35 GeV$. If $g_A$
grows larger, masses of $\kappa$ and $\sigma'$ will be pushed higher
too.
In both cases, with and without 't Hooft term's contribution, the
$\sigma'$  is problematic within the present scenario to be
identified with the physical $f_0(980)$ state, simply because the
former is too heavy. As will be seen in the discussion given later
in this paper, that in order to explain the large width of $\sigma$
and $\kappa$ in a dynamical approach, one needs large values of
$M_Q$ and $g_A$. The immediate consequence is that the ENJL model
would predict an unacceptably large vector meson spectrum and hence
fails to  give the correct description to the mass of $\rho$ and
$a_1$ mesons. The reason for this is because in ENJL model the
correlation between the parameters of scalar sector and the vector
seems to be too strong. This is not necessary for hadron physics --
in resonance chiral theory, for example, there is no such strong
correlations between the two sectors. Hence we will in the following
only  focus upon the scalar sector and ignore the problem in the
vector sector. The experience we are going to obtain is still
meaningful -- if not within the ENJL model itself -- in a more
general background, as in the resonance chiral theory.

In the scalar sector, as we mentioned above, the problem remains how
to identify the $f_0(980)$ resonance. A possible way to solve the
$\sigma'$ and the $f_0(980)$ problem is that since the bare state
$\sigma'$ is much heavier it may mix with $f_0(1370)$, etc. Without
instanton effects, the $\sigma$ and $\sigma'$ are ideally mixed and
the latter is $|\bar{s}s>$. When the instanton effects are taken
into account, $\sigma'$ may contain a sizable $|\bar{n}n>$ content
and hence may have a sizable mixing with the heavier scalar like
$f_0(1370)$, thus reducing to some extent its mass. On the other
side, one may identify the $\sigma'$ simply to the $f_0(1500)$
state, since the mass can be quite close to each other and the
latter is known to be mainly $\bar ss$ state. Then the $f_0(980)$
may be considered as a $\bar KK$ molecule~\rcite{locher,baru}.
Considering the complicated situation about $f_0(980)$, a convincing
explanation to $f_0(980)$ is out of the range of the present
discussion and remains to be explored in future.

 There are six
parameters ($M_Q$, $x$, $g_A$, $m_q$, $m_s$ and $\beta$) in the ENJL
model under investigation, and there are different ways to choose
physical parameters to be fit. For example, we can fit $f_\pi$,
$f_K$, $m_\pi$, $m_K$, $M_{th}$ (through
$M_{th}^2=m_\eta^2+m_{\eta'}^2-2m_K^2$), $m_\rho$ and the bare mass
of $\sigma$. In the old literature, the bare mass of $\sigma$ (and
also $a_0$ in the absence of anomaly) is typically 500 -- 600MeV
which is too small. Since roughly there is a mass relation in the
chiral limit, $M_\sigma\sim 2M_Q$~\cite{BRZ04}, it is difficult to
increase the bare mass of $\sigma$ within the ENJL model. Adding 't
Hooft interaction term will even further decrease the singlet
$\sigma$ mass. So the first thing to be noticed is that it is
somewhat unnatural to assign a sigma mass of order 1GeV in the ENJL
model. As can be seen from table 1, the $x$ parameter is quite
large, which is not natural in the cutoff effective lagrangian
approach. Another problem is that $f_V$ ($\simeq$ 0.1) can no longer
be fitted well to its experimental value ($\simeq 0.2$). Also the
current strange quark mass gets unnaturally large when $g_A$
increases. Furthermore, besides these problems,
 it is clear
  from table~\ref{table1}  that when $g_A$ gets
larger the mass of $\sigma'$ is enhanced and deviates more and more
from the narrow width state $f_0(980)$.
Barring this problem, setting $M_\sigma\sim 1$GeV, the bare mass of
the $\kappa$ resonance is an output which turns out to be $\sim
1.3$GeV and agrees within expectation.  Table~1 provides several fit
values.
\begin{table}
\caption{Experimental values and predictions of the ENJL model for
the various low energy parameters discussed in the text. All
dimensional quantities are in MeV except $m_{th}^2$ in GeV. }
\label{table1}
\begin{center}
\begin{tabular}{|c|c|c|c|c|c|c|c|c|}
\hline
    & exp. &    fit 1 & fit 2&fit 3&fit 4&fit 5 \\
    & value&         &      &     &     &     \\
\hline \hline
$ f_\pi $&92.4 &  92.6  &  92.5 &  92.3&92.0 &91.7 \\
$ f_K $& 112.0& 102.0  &  106.7& 112.4&118.9&136.1 \\
$ m_\pi  $& 137.3& 137.2 &  137.2& 137.2&137.3&137.3 \\
$ m_K  $&495.7&  495.6 & 495.7 & 495.6&495.6&495.4  \\
$ m_{th}^2 $&0.727& 0.645  & 0.44  &  0.31&0.23&0.16 \\
$ M_\sigma^\ddagger $&$\thicksim 930$&    856   & 869 & 881&892&903.8  \\
$ M_a $&  984.7 &    1039  & 1025&1016&1010&1004 \\
\hline
$ M_\kappa^\ddagger $&$\thicksim 1400$&   1227  & 1274&1330&1391&1458  \\
$ M_{\sigma'} $& 980&    1360  & 1456&1560&1669&1784 \\
\hline \hline
$x$  & &    0.175&  0.234 & 0.295&0.356&0.419  \\
$M_Q$& &    397.0&  395.3 & 394.1&393.3& 393.3 \\
$g_A$& & $0.5^*$&$0.6^*$ &$0.7^*$&$0.8^*$&$0.9^*$  \\
$m_q$  & &   4.6 &  6.9   & 9.6  & 12.7&16.1 \\
$m_s$  & & 114.  &  172.4 & 240.0&317.1&403.5 \\
$\beta$  & &  9.2  &  6.7   & 5.2  &  4.0&3.1 \\
\hline
\end{tabular}
\end{center}
(${}^{\ddagger}$) Corresponding to  bare masses discussed in the text. \\
($*$) Values of $g_A$ are fixed in the fits.
\end{table}
\subsection{Decays of Scalar Mesons}
A serious investigation of the scalar mass spectrum unavoidably
requires taking unitarization into account. But before doing that,
in this section we will discuss at tree level the decay widths of
light scalars, which can   be helpful, though very rough, in the
understanding of  strong interaction dynamics behind. For example,
if the decay width in a given channel in perturbation calculation is
small then we can judge that the interaction is not strong and the
difference between bare mass and pole mass is unimportant.  If on
the other hand the decay width is very large then one may claim that
the difference between bare mass and pole mass  ought to be large.
In the latter case one has to find more reliable method to handle
the strong interaction dynamics rather than calculating decay width
perturbatively.

\begin{figure}
\begin{center}
\begin{tabular}{lcr}
\epsfig{file=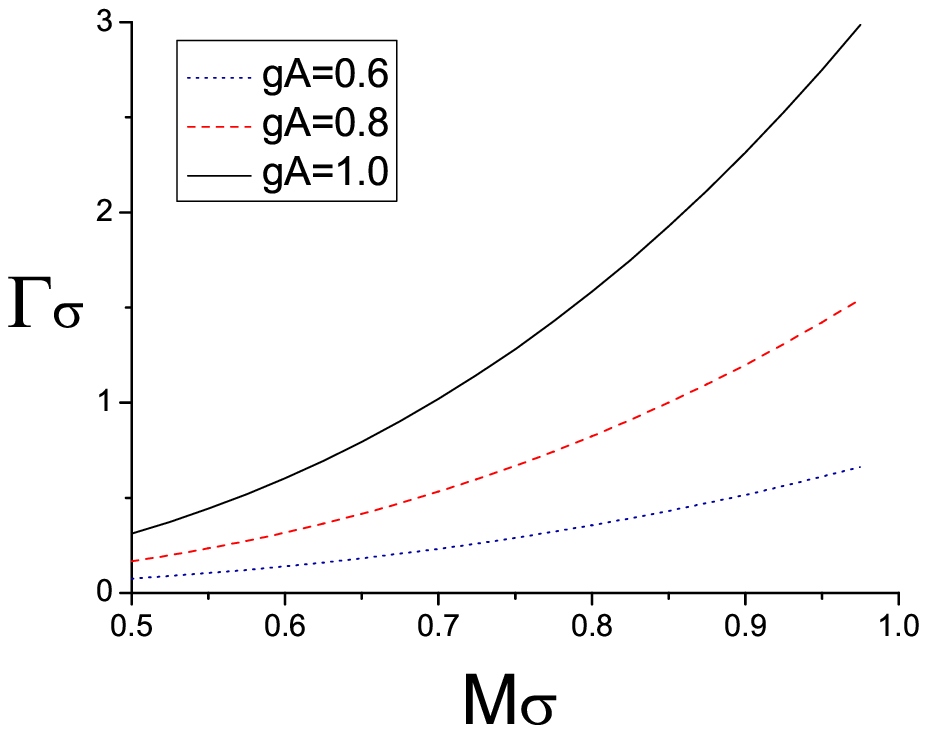,width=5.5cm}&\qquad\qquad\qquad&
\epsfig{file=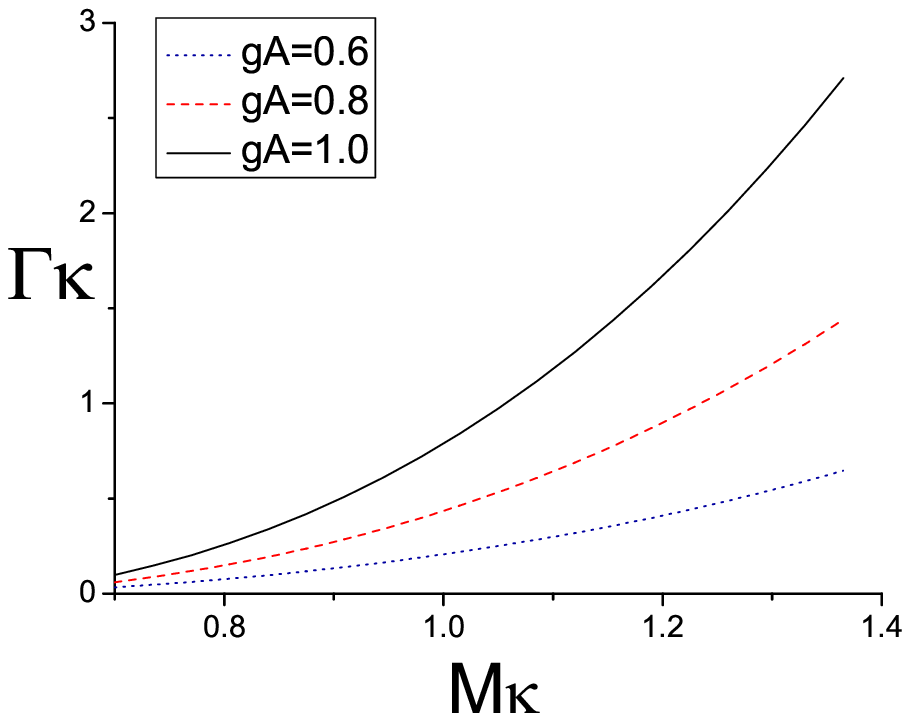,width=5.5cm}
\end{tabular}
\vspace*{8pt}
\caption{\label{fig1}%
The decay widths of $\sigma$ (left) and $\kappa$ (right)  as a
function of scalar mass $M_\sigma$ and $M_\kappa$, respectively; for
$g_A$=1.0, 0.8 and 0.6. The units are in GeV.}
\end{center}
\end{figure}

\begin{center}
\begin{figure}
{\epsfysize=5cm\epsfxsize=8cm\epsfbox{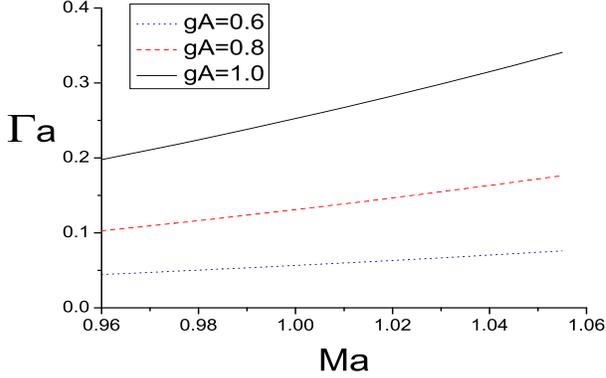}} \caption{The  decay
width  of  $(a_0\rightarrow \pi\eta_8)$ as a function of bare scalar
mass $M_{a_0}$  for values of $g_A$=1.0, 0.8 and 0.6 respectively. }
\label{fig2}
\end{figure}
\end{center}

We use the effective lagrangian to calculate the decay rates of a
scalar into two pseudoscalars, at tree level. The $\sigma$ decay
width is expressed below,
\begin{eqnarray}
\label{sigmatopipi} &&\Gamma _{\sigma ->\pi\pi} =3\Gamma _{\sigma
->\pi_0\pi_0} \nonumber \\
&& =\frac{g_A}{16\pi
M_{\sigma}f_\pi^2}\sqrt{1-\frac{4m_\pi^2}{M_\sigma^2}}
(g_A(M_\sigma^2-2m_\pi^2)+m_\pi^2)^2(\cos \theta
+\frac{\sin \theta}{\sqrt{2}})^2\ ,
\end{eqnarray}
where $\theta$ is the scalar meson mixing angel defined by
Eq.~(\ref{sigma mass}). When $\theta$ is equal to $\theta_{id}$, the
decay width is   maximal. From Fig.~\ref{fig1}, one realizes that in
order to explain the large discrepancy between $m_\sigma$ and
$M_\sigma$, $g_A$ should not be small, for otherwise the decay width
is small and the interaction will not be strong enough to develop a
big difference between $m_\sigma$ and $M_\sigma$.
Especially, from the Fig.~\ref{fig1}, we realize that the light
$M_\sigma \thicksim 500MeV$ and $M_\kappa \thicksim 700MeV$ can not
produce large widths. In the SU(3) limit, we have $g_{\sigma
\pi\pi} =g_{\kappa K\pi}$ and the width is proportional to $g_A^3$.
The only possibility in both cases to get a large width is to
increase the bare mass parameters. We also plot the decay width of
$a_0\rightarrow \pi\eta_8$ in Fig.~\ref{fig2}. The decay width of
$a_0$ is much smaller comparing with that of $\sigma$ and $\kappa$
simply because of SU(3) symmetry. See Fig.~\ref{fig2} for
illustration. Therefore, as revealed by Figs.~1 and 2, the ENJL
model does provide a possibility in its parameter space to explain
the observed scalar spectrum and the vastly different widths
simultaneously, at least qualitatively.
\section{Pole masses of scalar resonances in the ENJL model}
\subsection{The K matrix unitarization and the pole
positions}\label{Kmatrix}
Certain unitarization approximation is necessary when a large width
is involved. The unitarization method has been applied to
(resonance) chiral perturbation theory amplitudes, and also to
linear sigma models in the literature (see for example
Ref.~\rcite{Black,Pich}). To our knowledge, this paper is the first
attempt to apply unitarization to ENJL amplitudes. The scattering
amplitudes for two pseudoscalars to two pseudoscalars are easily
obtainable at tree level in the ENJL model, The single channel
K--matrix unitarization is the following:
 \be\label{K1}
  T=\frac{T^{\mathrm{tree}}}{1-i\rho T^{\mathrm{tree}}}\ .
   \ee
We use the $K$ Matrix amplitude determined from ENJL model to search
for pole positions of scalars, which are not found in the previous
literature. The results, corresponding to several choices of
$g_A=0.6, 0.7$ and 0.8, are listed in table~\ref{table3}.   In the
unitarized amplitudes there are actually quite a few poles in each
channel, on both sheets. Nevertheless   in each channel there is
only one pole that falls on the real axis in the large $N_c$ limit
which is just the input pole in the lagrangian.\footnote{The
$\sigma'$ is very heavy and lies far above the $\pi\pi$ elastic
unitarity region and hence we do not attempt to make any discussion
based on the unitarized amplitude.} However, the results listed in
table~\ref{table3} should not be understood as accurate in any
sense. On the contrary, it is known that the $K$ matrix results are
crude for derivative coupling theories~\cite{guo}. The results given
in table~\ref{table3} only provide a qualitative guide to the
underlining dynamics: when $g_A$ is small the coupling strength
between $\sigma$ and $\pi\pi$ is small and the width of $\sigma$ is
also small. The mass of the $\sigma$ found from the unitarized
amplitude is therefore very close to its input value. However, when
$g_A$ increases up to, for example, 0.8, the width of $\sigma$
becomes large, and the pole mass $m_\sigma$ becomes totally
different from the input bare mass, $M_\sigma$.
\begin{table}
\caption{Scalar pole masses  } \label{table3}
\begin{center}
\begin{tabular}{|c|c|c|c|  }
\hline
    & fit 2 ($g_A=0.6$)&  fit 3 ($g_A=0.7$)&  fit 4  ($g_A=0.8$) \\
    \hline \hline
$\pi \pi\rightarrow \pi \pi$ & $985\pm133i$   & $1161\pm229i$ & $544\pm306i$  \\
$\pi K\rightarrow \pi K$ & $1423\pm153i$  & $1629\pm202i$ & $801\pm360i$ $^{*}$ \\
$ \pi \eta\rightarrow \pi \eta $ & $1030\pm31i$  & $1016\pm45i$ & $1000\pm60i$ \\
\hline \hline
\end{tabular}
\end{center}
$^*$: The $N_c$ trajectory is marginal.
\end{table}

\subsection{Pole trajectories with respect to the variation of $N_c$}
As stated in last section that, all poles listed in
table~\ref{table3} fall on the real axis in the large $N_c$
limit.\footnote{For $g_A=0.8$ the kappa pole trajectory is actually
marginal, the $\kappa$ pole will fall on the real axis when further
increasing $g_A$.} However, there are other poles on the second
sheet and it is checked that they all go to $\infty$ on the complex
$s$ plane when $N_c\to \infty$. Hence these states are dynamically
generated. As discussed in section~\ref{Kmatrix}, for small values
of $g_A$ (for example, fit 2 and fit 3) the $\sigma$ pole from the
ENJL lagrangian has a rather small width and a large mass around
1GeV (as an input), but it is observed that in such a case there
still exists a light and broad dynamical pole which disappears when
$N_c\to\infty$. This pole, being dynamical, is certainly not the
$\sigma$ pole responsible for chiral symmetry breaking in the ENJL
model, since the latter is well monitored and falls on the real axis
in the large $N_c$ limit. One may even further ask the question
whether the experimentally observed $f_0(600)$ is the $\sigma$
responsible for chiral symmetry breaking (In the present situation
corresponding to the light and broad resonance when $g_A$ is large
as in fit 4), or a dynamically generated light and broad resonance,
which is not the $\sigma$, when $g_A$ is small. To understand why
there appears a `dynamical pole' we recall that in general the tree
level IJ=00 channel $\pi\pi$ elastic scattering amplitude, in the
chiral limit, may be written as
\begin{eqnarray}\label{dynLsigmaM}
&&T_{C.A.}=\frac{s}{16\pi f_\pi^2}\, ,\nonumber\\
&&T_4=\frac{s^2}{24\pi
f_\pi^4}\triangle\,  ,        \,\,(\triangle=22{\tilde L}_1+14{\tilde L}_2+11{\tilde L}_3)\nonumber\\
&&T_S=\frac{c_d^2}{16\pi
f_\pi^4}(s-2M_\sigma^2+\frac{3s^2}{M_\sigma^2-s}+\frac{2M_\sigma^4}{s}\ln[1+\frac{s}{M_\sigma^2}]),\,
\,\,\end{eqnarray} where $c_d\sim O(\sqrt{N_c})$ and $\tilde{L_i}$
are obtained by integrating out all resonance fields  except
scalars. The above expressions generally depict  resonance chiral
theory amplitudes including the ENJL model. The pole position in the
chiral limit is determined by the equation
\begin{equation}\label{dynLsigmaM'}
1+i (T_{CA}+T_4+T_S)=0.
\end{equation}
In Eq.~(\ref{dynLsigmaM'}) if we set $T_4$ and $T_s$ vanishing, we
get the `current algebra sigma' pole as already discussed in
Ref.~\cite{guo}. The $N_c$ dependence of the `current algebra sigma'
pole position is $\sqrt{s_{pole}}\sim \sqrt{N_c}$ and is ruled out
through the study of Ref.~\cite{guo,Guo07}. If setting $T_4=0$ and
$\sqrt{2}c_d/f_\pi=1$ in Eq.~(\ref{dynLsigmaM}) we recover the
linear $\sigma$ model amplitude.\footnote{In ENJL model we have
approximately $\sqrt{2}c_d/f_\pi\simeq g_A^{3/2}$. In resonance
chiral  theory it is found that $\sqrt{2}c_d/f_\pi\simeq
0.53$~\cite{ecker} which corresponds to $g_A\simeq 0.65$ here.}  In
such a case the $\sigma$ resonance is light and broad when the bare
mass of $\sigma$ is around 1GeV. In general, however, if we neglect
the logarithm term in Eq.~(\ref{dynLsigmaM'}), which is suppressed
when $s$ is large, it is not difficult to show that on the second
sheet there exists, except the stable pole in the large $N_c$ limit,
another pole with the property $s_{pole}\sim\sqrt{N_c}$ on the
second sheet of complex s plane. Notice that dynamical pole obtained
from Eq.~(\ref{dynLsigmaM'}) contains a different $N_c$ behavior
comparing with the `current algebra $\sigma$': the latter behaves as
$\sqrt{s_{pole}}\sim\sqrt{N_c}$. The different $N_c$ dependence of
the pole trajectory actually reminds us that the property of the so
called `dynamical' pole can be highly (unitarization) model
dependent.

To prove the illegality of the light and broad dynamical pole
generated from simple $K$ matrix unitarization of the tree level
ENJL amplitude when $g_A$ is small, we make use of the low energy
matching method developed in Ref.~\rcite{Xiaozg} (see also
Ref.~\cite{Guo07}). For $\pi\pi$ scattering $S$ matrix poles (on the
second sheet) obey one relation:
\begin{eqnarray}
\label{matching1} &&\sum_R \frac{G_R}{M_R^2-4m_\pi^2}\sim
O(N_c^{-1})\,,
\end{eqnarray}
where $M_R^2$ and $G_R$ are functions of the pole mass $z_0$ of
resonance $R$~\rcite{Zheng1}:
\begin{eqnarray}
&&M_R^2(z_0)=\mathrm{ Re}(z_0)+\frac{\mathrm{
Im}(z_0)\mathrm{Im}[z_0 \rho(z_0)]}{\mathrm{Re}[z_0
\rho(z_0)]},\\
&&G_R(z_0)=\frac{\mathrm{Im}(z_0)}{\mathrm{Re}[z_0\rho(z_0)]},\,  \,\\
&&\rho(z_0)=\sqrt{1-4m_\pi^2/z_0}\ .
\end{eqnarray}
No matter where on the second sheet does the pole locate, one
always has ${G_R}/{(M_R^2-4m_\pi^2)}>0$. The pole solution of
Eq.~(\ref{dynLsigmaM}) with the property $s_{pole}\sim\sqrt{N_c}$
corresponds to $M_R^2\propto\textit{O}(\sqrt{N_c})$ and
$G_R\propto \textit{O}(1)$. Its contribution to the $l.h.s$ of
Eq.~(\ref{matching1}) is $\textit{O}(1/\sqrt{N_c})$, meanwhile,
the right-hand side is $\textit{O}(1/N_c)$. The only possibility
to satisfy Eq.~(\ref{matching1}) is that the contribution of a
such `dynamical' pole is canceled by a spurious pole on the
physical sheet, whose contribution is also of order of
$\textit{O}(1/\sqrt{N_c})$. This is just the case what we found
from solutions of Eq.~(\ref{dynLsigmaM}). Hence we demonstrate
that the dynamical light and broad pole in the ENJL model,
generated in the present simple $K$ matrix unitarization,  is
spurious.

We can also check the [1,1] Pad\'e amplitude. The pole position of
the [1,1] Pad\'e amplitude is determined by the equation, in the
chiral limit,
\begin{eqnarray}\label{Pade}
&&T_2-(T_4+T_s-i|T_2|^2)=0\nonumber\\
&&\sim \frac{s}{16\pi f_\pi^2}-\frac{s^2}{24\pi
f_\pi^4}\Delta-\frac{c_d^2}{16\pi
f_\pi^4}(s-2M_\sigma^2+\frac{3s^2}{M_\sigma^2-s})+i
(\frac{s}{16\pi f_\pi^2})^2=0. \end{eqnarray} As before we neglect
the logarithm term in above. It is straightforward to show that if
 no accidental cancelation occurs, when $N_c\to\infty$, there
exist two poles on the second sheet of complex s plane, one is on
the real axis and the other remains on the complex s plane:
$s_{pole}\propto \textit{O}(1)$. At the same time, a spurious pole
on the first sheet will be found, which is also $s_{pole}\propto
\textit{O}(1)$. The latter exactly cancels the second sheet pole to
meet the $N_c$ order of  the $l.h.s$ of Eq.~(\ref{matching1}). Hence
the dynamical pole with $s_{pole}\propto \textit{O}(1)$ found from
the amplitude~(\ref{Pade}) is also a spurious one. Therefore the
situation as described by Eq.~(\ref{Pade}) is quite different from
the Pad\'e amplitudes constructed from pure chiral perturbation
theory~\cite{padechiral,pelaez06}. The latter is obtained by further
integrating out the explicit scalar degree of freedom. There one
does find that the dynamical pole falls on the real axis in the
$N_c\to \infty $ limit.

\section{Discussions and Conclusions }
In this paper we discuss the possibility whether one can understand
the light and broad $\sigma$ and $\kappa$, together with the narrow
$a_0(980)$ and $f_0(980)$ in a same SU(3) nonet, in the ENJL model.
We find that the ENJL model is quite reluctant for this picture. One
difficulty is that the $\sigma'$ resonance is simply too heavy to be
identified as the $f_0(980)$ meson. One has to call for other
mechanisms for the rescue. For example, the mixing with $f_0(1370)$
and/or $f_0(1500)$; or that $f_0(980)$ is simply a $\bar KK$
molecular state~\cite{baru,markushin}. Beside this difficulty,
however, the ENJL model can give a rough but unified description to
the light and broad $\sigma$, $\kappa$ and the narrow $a_0(980)$.
For sufficiently large $g_A$ and an input bare $\sigma$ mass around
1GeV, a simple unitarization approximation generates a light and
broad $\sigma$ resonance. The difference comes from the fact that
the $\sigma$ couples very strongly to $\pi\pi$ continuum, hence its
pole location is severely distorted.  The price paid for this
picture is that the $g_A$ and $M_Q$ parameter have to be unnaturally
large. As a consequence, the ENJL model is no longer valid for
describing the vector meson spectrum. However, if we disregard the
constraints among parameters of ENJL model, the above picture can be
realized without any foreseeable difficulty in general.

We also discussed the fate of dynamical poles generated from the
simple $K$ matrix unitarized $\pi\pi$ scattering amplitude, when
$g_A$ is small. It was confusing to notice that, in such a case,
there still exists a light and broad dynamical pole which might be
identified as the observed $f_0(600)$ resonance, besides the input
heavy (and narrow) $\sigma$ pole. However, we find that this
dynamical pole maintains a wrongful $N_c$ behavior which has to be
canceled by an accompanying  first sheet pole, hence violating
analyticity and should be spurious. The lesson we learn from this
study is that one has to be extremely cautious when trying to give a
physical meaning to a dynamically generated resonance pole from a
unitarized amplitude. The property of the latter can be highly model
dependent. Finally further efforts have to be made in order to
 generate  successfully the light and broad scalar spectrum from a
general resonance lagrangian containing both the scalar and the
vector sectors.

{\bf Acknowledgement:} We would like to thank  Zhi-Hui Guo and
Juan Jose Sanz-Cillero for helpful discussions. This work is
supported in part by National Natural Science Foundation of China
under contract number
 10575002, 
and  10421503.
\newpage
\appendix
\section{Effective Couplings in the Effective Lagrangian}

We list in the following parameters of effective meson lagrangian
obtained from ENJL model. The following expressions are found in
agreement with those given in Ref.~\rcite{Bijnens}. In the
calculation, only those regularization scheme independent terms
(leading terms in cutoff  dependence) are kept.
\begin{align*}\label{append1}
&f^{2}   =\frac{N_{c}}{16\pi^{2}}\Gamma(0,x)4M_{Q}^{2}g_{A},\,\,%
B=\frac{N_{c}}{16\pi^{2}}\Gamma(-1,x)\frac{4M_{Q}^{3}}{f^{2}}=\frac
{\Gamma_{-1}\cdot M_{Q}}{\Gamma_{0}\cdot g_{A}},\\
&M_V^2 = {3 \over 2} {\Lambda_{\chi}^2 \over G_V(\Lambda_{\chi}^2)}
{1 \over \Gamma(0,x)}=6 M_Q^2\frac{g_A}{1-g_A}, \\
&M_A^2= (M_V^2 + 6M_Q^2)/(1- {\Gamma(1,x) \over
\Gamma(0,x)}),\\
&  f_V = \sqrt 2 \lambda_V\,\,\,\,\, , \,\,\,\,\,f_A = \sqrt 2 g_A \lambda_A, \\
& g_V = {N_c \over 16\pi^2} {1 \over \lambda_V} {\sqrt 2 \over 6}
\left[(1-g_A^2) \Gamma(0,x) + 2 g_A^2
\Gamma(1,x) \right],\\
&\lambda_S^2 = {N_c \over 16\pi^2} {2 \over 3} \left[3\Gamma(0,x) -
2\Gamma(1,x) \right],
\\
&M_S^2 = {N_c \over 16\pi^2} {8M_Q^2  \over \lambda_S^2}
\Gamma(0,x),\\
 &  c_m = {N_c \over 16\pi^2} {M_Q \over \lambda_S} \rho
\left[\Gamma(-1,x) - 2 \Gamma(0,x) \right],\\
& c_d = {N_c \over 16\pi^2} {M_Q \over \lambda_S} 2
g_A^2\left[\Gamma(0,x) -\Gamma(1,x)\right],
\end{align*}

\begin{align*}
&  \tilde{L}_{1}=\frac{1}{2}\frac{N_{c}}{16\pi^{2}}\left[
\Gamma(0,x)\frac
{(1-g_{A}^{2})^{2}}{24}\right]  ,\\
&  \tilde{L}_{2}=\frac{N_{c}}{16\pi^{2}}\left[  \Gamma(0,x)\frac{(1-g_{A}^{2})^{2}%
}{24}\right]  ,\\
&  \tilde{L}_{3}=\frac{N_{c}}{16\pi^{2}}\left[  -\Gamma(0,x)\frac{(1-g_{A}^{2})^{2}%
}{8}\right] , \\
&  \tilde{L}_{5}=\frac{N_{c}}{16\pi^{2}}\left[ \Gamma(0,x)\right]
g_{A}^{2}M_{Q}\frac{1}{2B},\\
&  \tilde{L}_{7}=\frac{N_{c}}{16\pi^{2}}\left[  -\Gamma(0,x)g_{A}M_{Q}\frac{1}%
{12B}\right]  ,\\
&  \tilde{L}_{8}=\frac{N_{c}}{16\pi^{2}}\left[  \Gamma(-1,x)\frac{M_{Q}^{2}}{8B^{2}%
}\right]  ,\\
&  \tilde{H}_{2}=\frac{N_{c}}{16\pi^{2}}\left[  \Gamma(-1,x)\frac{M_{Q}^{2}}{4B^{2}%
}\right].
\end{align*}

In the following we list expressions of scalar mass and pseudoscalar
decay constants. Some of them are not found in the previous
literature.
\begin{eqnarray}
M_a^2&=&M_{a_0}^2+\frac{1}{\lambda_s^2}(4\beta
M_Q+\frac{2}{\sqrt{3}}\beta v_0-\frac{2}{\sqrt{6}}\beta v_8),\nonumber\\
M_\kappa^2&=&M_{\kappa_0}^2+\frac{1}{\lambda_s^2}(4\beta
M_Q+\frac{2}{\sqrt{3}}\beta v_0+\frac{2}{\sqrt{6}}\beta v_8),\nonumber\\
M_{a_0}^2&=&\frac{N_C\Gamma[0,x]}{16\pi^2
\lambda_s^2}4M_Q^2[2+\frac{3}{M_Q}(m_d+m_u+\frac{2}{\sqrt{3}}v_0+\frac{2}{\sqrt{6}}v_8)],\nonumber\\
M_{\kappa_0}^2&=&\frac{N_C\Gamma[0,x]}{16\pi^2
\lambda_s^2}4M_Q^2[2+\frac{3}{M_Q}(m_s+m_u+\frac{2}{\sqrt{3}}v_0-\frac{1}{\sqrt{6}}v_8)],\nonumber\\
M_{00}^2&=&4M_Q^2+4M_Q(m_u+m_d+m_s)+\frac{12}{\sqrt{3}}v_0
M_Q\nonumber\\
&-&\beta\frac{M_Q+\frac{2}{\sqrt{3}}v_0}{\frac{N_C}{16\pi^2}\Gamma[0,x]},\nonumber\\
M_{88}^2&=&4M_Q^2+2M_Q(m_u+m_d+4m_s)+\frac{12}{\sqrt{3}}v_0
M_Q-2\sqrt{6}v_8M_Q\nonumber\\
&+&\beta\frac{2M_Q+\frac{1}{\sqrt{3}}v_0+\frac{2}{\sqrt{6}}v_8}{\frac{N_C}{16\pi^2}\Gamma[0,x]},\nonumber\\
M_{08}^2&=&4\sqrt{2}(m_d+m_u-2m_s)M_Q\nonumber\\
&+&2v_8(4\sqrt{3}M_Q+\frac{\beta}{\sqrt{3}\frac{N_C}{16\pi^2}\Gamma[0,x]}),\nonumber\\
M^2_\sigma&=&\frac{1}{2}\left[M_{00}^2+M_{88}^2-
                           \sqrt{(M_{00}^2-M_{88}^2)^2+(M_{08}^2)^2}\right], \nonumber \\
M^2_{\sigma'}&=&\frac{1}{2}\left[M_{00}^2+M_{88}^2+
                           \sqrt{(M_{00}^2-M_{88}^2)^2+(M_{08}^2)^2}\right].\nonumber
\end{eqnarray}

\begin{eqnarray}
f_\pi^2&=&\frac{N_C}{4\pi^2}\Gamma[0,x]g_{1A} M_Q^2+\frac{N_C}{2\pi^2}\Gamma[0,x]g_AM_Q(\frac{v_0}{\sqrt{3}}+\frac{v_8}{\sqrt{6}}+m_q),\nonumber\\
f_K^2&=&\frac{N_C}{8\pi^2}M_Q^2(g_{1A}+g_{2A})+\frac{N_C}{4\pi^2}\Gamma[0,x]g_A M_Q(\frac{2}{\sqrt{3}}v_0-\frac{1}{\sqrt{6}}v_8+m_s+m_q),\nonumber\\
m_\pi^2&=&2M_Q m_q\frac{\Gamma[-1,x]}{g_A\Gamma[0,x]},\nonumber\\
m_K^2&=&M_Q (m_q+m_s)\frac{\Gamma[-1,x]}{g_A\Gamma[0,x]},\nonumber\\
g_{1A}&=&g_A(1-\frac{1}{3M_Q}(6m_q+2\sqrt{3}v_0+\sqrt{6}v_8)(1-g_A)),\nonumber\\
g_{2A}&=&g_A(1-\frac{1}{3M_Q}(6m_s+2\sqrt{3}v_0-2\sqrt{6}v_8)(1-g_A)),\nonumber\\
v_0&=&\frac{2(2m_q+m_s)}{\sqrt{3}(4-\frac{\beta}{M_Q\frac{N_C}{16\pi^2}\Gamma[0,x]})}\frac{\Gamma[-1,x]}{\Gamma[0,x]},\nonumber\\
v_8&=&\frac{2\sqrt{2}(m_q-m_s)}{\sqrt{3}(4+\frac{2\beta}{M_Q\frac{N_C}{16\pi^2}\Gamma[0,x]})}\frac{\Gamma[-1,x]}{\Gamma[0,x]}.\nonumber
\end{eqnarray}
$v_0$ and $v_8$ are the vacuum expectation values of the SU(3)
singlet field $\sigma_0$ and the octet field $\sigma_8$ of the
lowest order of the current quark mass in the broken phase. $f_\pi$
and $f_K$ will get the tadpole's contribution.

\end{document}